\documentclass[12pt]{article}
\usepackage[dvips]{graphicx}
\usepackage{epsfig, comment}
\usepackage{cite}
\usepackage{latexsym,amsmath,amsfonts,amssymb}
\usepackage{mathtools} 
\usepackage{subcaption}
\usepackage{slashed} 
\usepackage{blkarray}
\usepackage[all]{xy} 
\usepackage[makeroom]{cancel}
\usepackage[latin1]{inputenc}
\usepackage[american]{babel}
\usepackage{bbm}
\usepackage{color}
\usepackage[unicode]{hyperref}
\definecolor{rosy}{RGB}{230,235,252}
\definecolor{myframetitle}{RGB}{90,89,170}
\definecolor{myblocktitle}{RGB}{140,185,249}
\definecolor{mytitle}{RGB}{10,80,26}

\definecolor{darkgreen}{RGB}{27,130,45}
\definecolor{darkblue}{rgb}{0,0,0.3}
\definecolor{darkred}{rgb}{0.7,0,0}

\definecolor{light gray}{RGB}{220,220,220}
\definecolor{dark purple}{RGB}{108,0,217}
\definecolor{pink}{RGB}{190,20,100}
\definecolor{orang}{RGB}{193,63,0}
\definecolor{green}{RGB}{11,98,17}
\definecolor{darkpink}{RGB}{153,0,76}
\definecolor{bluegreen}{RGB}{0,102,102}
\definecolor{greenlagan}{RGB}{0,102,0}
\definecolor{redgreen}{RGB}{102,102,0}
\definecolor{Redgreen}{RGB}{153,76,0}
\definecolor{vividviolet}{rgb}{0.62, 0.0, 1.0}
\definecolor{amaranth}{rgb}{0.9, 0.17, 0.31}
\definecolor{palatinateblue}{rgb}{0.15, 0.23, 0.89}
\definecolor{brightpink}{rgb}{1.0, 0.0, 0.5}
\definecolor{cornflowerblue}{rgb}{0.39, 0.58, 0.93}
\definecolor{deepcarminepink}{rgb}{0.94, 0.19, 0.22}
\definecolor{radicalred}{rgb}{1.0, 0.21, 0.37}

\hypersetup{ linktoc=all,
    colorlinks, linkcolor={palatinateblue},
    citecolor={brightpink}, urlcolor={amaranth}
}

\graphicspath{{Images/}}

\def\6{\partial}

\newcommand{\be}{\begin{equation}}
\newcommand{\ee}{\end{equation}}
\newcommand{\beq}{\begin{equation}}
\newcommand{\eeq}{\end{equation}}
\newcommand{\bea}{\begin{eqnarray}}
\newcommand{\eea}{\end{eqnarray}}
\newcommand{\nn}{\nonumber \\}

\newcommand{\ba}{\begin{eqnarray}}
\newcommand{\ea}{\end{eqnarray}}

\newcommand{\beqs}{\begin{eqnarray}}
\newcommand{\eeqs}{\end{eqnarray}}
\newcommand{\bal}{\begin{aligned}}
\newcommand{\eal}{\end{aligned}}

\def\lbldef#1#2{\expandafter\gdef\csname #1\endcsname {#2}}

\def\href#1#2{#2}

\newcommand{\ber}{\begin{eqnarray}}
\newcommand{\eer}{\end{eqnarray}}

\newcommand{\beqar}{\begin{eqnarray}}

\newcommand{\eeqar}{\end{eqnarray}}

\newcommand{\dsl}
   {\kern.06em\hbox{\raise.15ex\hbox{$/$}\kern-.56em\hbox{$\partial$}}}

\newcommand{\eeqarr}{\end{eqnarray}}
\newcommand{\ZZ}{{\rm \kern 0.275em Z \kern -0.92em Z}\;}

\def\CC{{\mathchoice
{\rm C\mkern-8mu\vrule height1.45ex depth-.05ex
width.05em\mkern9mu\kern-.05em}
{\rm C\mkern-8mu\vrule height1.45ex depth-.05ex
width.05em\mkern9mu\kern-.05em}
{\rm C\mkern-8mu\vrule height1ex depth-.07ex
width.035em\mkern9mu\kern-.035em}
{\rm C\mkern-8mu\vrule height.65ex depth-.1ex
width.025em\mkern8mu\kern-.025em}}}

\def\RR{{\rm I\kern-1.6pt {\rm R}}}

\def\ZZ{{\rm Z}\kern-3.8pt {\rm Z} \kern2pt}
\def\IB{\relax{\rm I\kern-.18em B}}
\def\ID{\relax{\rm I\kern-.18em D}}
\def\II{\relax{\rm I\kern-.18em I}}
\def\IP{\relax{\rm I\kern-.18em P}}

\newcommand{\bear}{\begin{eqnarray}}
\newcommand{\eear}{\end{eqnarray}}

\def\6{\partial}

\newfont{\namefont}{cmr10}
\newfont{\addfont}{cmti7 scaled 1440}
\newfont{\boldmathfont}{cmbx10}
\newfont{\headfontb}{cmbx10 scaled 1728}

\allowdisplaybreaks[1]

\begin{document}

\begin{titlepage}

\vfill
\begin{flushright}
\end{flushright}


\begin{center}

   \baselineskip=16pt
   {\Large \bf On Problems with Cosmography in Cosmic Dark Ages}
   \vskip 1cm
   A. Banerjee$^{a,b}$, E. \'O Colg\'ain$^{a, c}$, M. Sasaki$^{d, e, f}$, M. M. Sheikh-Jabbari$^{g}$, T. Yang$^a$
          \vskip .6cm
             \begin{small}
               \textit{$^a$ Asia Pacific Center for Theoretical Physics, Postech, Pohang 37673, Korea}
               
               \vspace{3mm} 
               \textit{$^b$ Okinawa Institute of Science and Technology, 1919-1 Tancha, Onna-son, Okinawa 904-0495, Japan }
               \vspace{3mm}
               
               \textit{$^c$ Department of Physics, Sogang University, Seoul 121-742, Korea}
               
               \vspace{3mm} 
                \textit{$^d$ Kavli Institute for the Physics and Mathematics of the Universe (WPI), \\ UTIAS, The University of Tokyo, Chiba 277-8583, Japan}
                
                \vspace{3mm} 
                \textit{$^e$ CGP, Yukawa Institute for Theoretical Physics, Kyoto University, Kyoto 606-8502, Japan}
                
                \vspace{3mm} 
                \textit{$^f$ LeCosPA, National Taiwan University, Taipei 10617, Taiwan}

		\vspace{3mm}

               \textit{$^g$ School of Physics, Institute for Research in Fundamental Sciences (IPM), \\
P.O.Box 19395-5531, Tehran, Iran}

             \end{small}
\end{center}

\begin{center} \textbf{Abstract}\end{center} \begin{quote}
Quasars show considerable promise as standard candles in a high-redshift window beyond Type Ia supernovae. Recently, Risaliti, Lusso \& collaborators \cite{Risaliti:2018reu, Lusso:2019akb, Lusso:2020pdb} have succeeded in producing a high redshift  Hubble diagram ($ z 
\lesssim 7$) that supports ``a trend whereby the Hubble diagram of quasars is well reproduced by the standard flat $\Lambda$CDM model up to $z \sim 1.5-2$, but strong deviations emerge at higher redshifts". This conclusion hinges upon a log polynomial expansion for the luminosity distance. In this note, we demonstrate that this expansion (or ``improvements'' thereof) typically can only be trusted up to $z \sim 1.5-2$. As a result, a breakdown in the validity of the expansion may be misinterpreted as a (phantom) deviation from flat $\Lambda$CDM. We further illustrate the problem through mock data examples. 

\end{quote} \vfill

\end{titlepage}

\section*{Background}
Quasars (QSOs), assuming they can be standardised, offer the potential to unlock a redshift range beyond the reach of Type Ia supernovae {that can stretch into the cosmic dark ages} \cite{Risaliti:2018reu, Lusso:2019akb, Lusso:2020pdb}. Ideally, one would like to analyse high-redshift data in as ``model independent" a methodology as possible. Arguably, the simplest technique in this class is ``cosmography", or more simply put, Taylor expansions about $z=0$ \cite{Visser-TE}. Unfortunately,  19$^{\textrm{th}}$ century mathematics \cite{CH} (as explained in  \cite{Cattoen:2007sk}) precludes Taylor expansions in a Friedmann-Lema\^itre-Robertson-Walker (FLRW) cosmology beyond redshift $z \sim 1$ {(see \cite{Colgain:2021ngq} for more discussions)}. In view of the recent claims \cite{Risaliti:2018reu, Lusso:2019akb, Lusso:2020pdb}, it is timely to remind the cosmology community that results in mathematics date extremely well. 

Within this context, the claims of $\sim 4 \sigma$ deviations from flat $\Lambda$CDM \cite{Risaliti:2018reu, Lusso:2019akb} raise an immediate red flag. The problem here is that while $\sim 4 \sigma$ may be an accurate reflection of the discrepancy of the data with the log polynomial employed extensively in \cite{Risaliti:2018reu, Lusso:2019akb, Lusso:2020pdb}, the log polynomial itself is not anchored to flat $\Lambda$CDM beyond $z\sim 1$. Moreover, the discrepancy between the log polynomial expansion and flat $\Lambda$CDM depends not only on cosmological parameters, i. e. matter density $\Omega_{m}$, but also the degree of the expansion. In essence, the ``yardstick" that one is using to compare the data to flat $\Lambda$CDM is a variable one, and for this reason, it is impossible to quantify any tension: {a ``$\sim 4 \sigma$ deviation from the flat $\Lambda$CDM model" is merely a mirage}. Note, we are not saying that the tension is not real, simply that it cannot be substantiated by the analysis presented in \cite{Risaliti:2018reu, Lusso:2019akb, Lusso:2020pdb}. This may be rectified elsewhere in the literature, e. g. \cite{Yang:2019vgk}.

\section*{Analysis}
In \cite{Risaliti:2018reu, Lusso:2019akb, Lusso:2020pdb}, the luminosity distance is expanded in terms of log polynomials, 
\be
\label{dl}
d^{\textrm{log poly}}_{L} (z) = k \ln (10) \frac{c}{H_0} \sum_{n=1}^4 a_n [\log_{10} (1+z)]^n + \dots 
\ee
where $k$ is degenerate with $H_0$, so it can be set to unity, $a_1 = 1$, and $\dots$ denote truncated terms. Relative to  \cite{Risaliti:2018reu}, which studies a third order polynomial ($n=3$), \cite{Lusso:2019akb} considers the additional $a_4$ term, while \cite{Lusso:2020pdb} adds an $a_5$ term (without exploring it). To connect the $a_i$ parameters of  $d^{\textrm{log poly}}_{L} (z)$ to those of flat $\Lambda$CDM, the following identities are employed \cite{Lusso:2019akb}: 
\bea
\label{equation}
a_2 &=& \ln (10) \left( \frac{3}{2} - \frac{3}{4} \Omega_m \right), \quad a_3 = \ln^2(10) \left( \frac{9}{8} \Omega_m^2 - 2 \Omega_m + \frac{7}{6} \right), \nn
a_4 &=& \ln^3(10) \left( - \frac{135}{64} \Omega_m^3 + \frac{9}{2} \Omega_m^2 - \frac{47}{16} \Omega_m + \frac{5}{8} \right), \nn
a_5 &=& \ln^4(10) \left( \frac{31}{120} - \frac{25}{8} \Omega_m + \frac{315}{32} \Omega_m^2 - \frac{729}{64} \Omega_m^3 + \frac{567}{128} \Omega_m^4 \right), 
\eea
where we have added the last relation. Note, these relations are based on Taylor expansion in $z$ about $z=0$ and by construction they guarantee that the exact $d_{L}^{\Lambda \textrm{CDM}}(z)$ of flat $\Lambda$CDM agrees with $d^{\textrm{log poly}}_{L} (z)$ only at low redshift: nothing is guaranteed at higher $z$. 

\begin{figure}[htb]
\centering
\begin{subfigure}{.35\textwidth}
  \includegraphics[width=55mm]{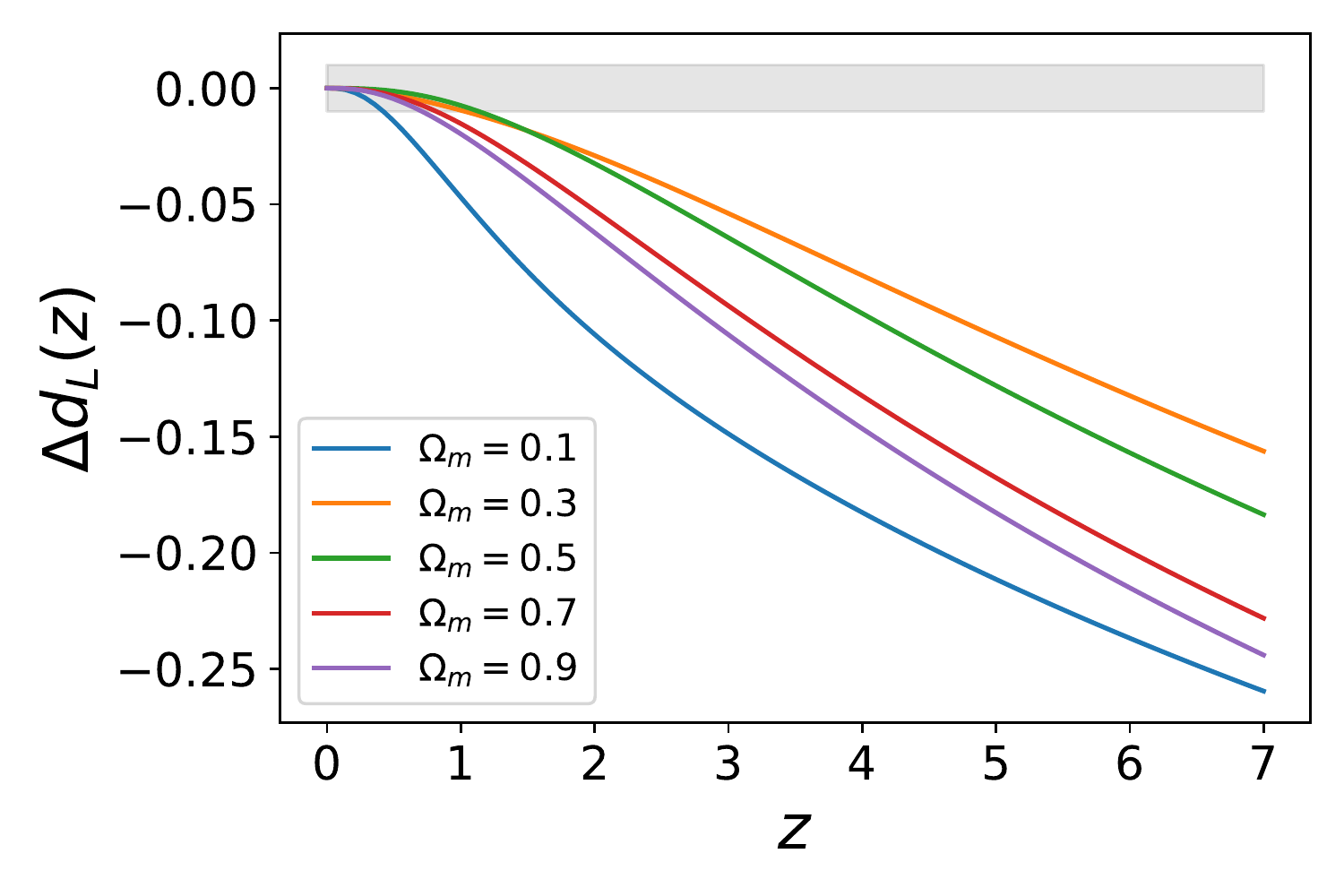}
  \caption{}
\end{subfigure}%
\begin{subfigure}{.35\textwidth}
  \includegraphics[width=55mm]{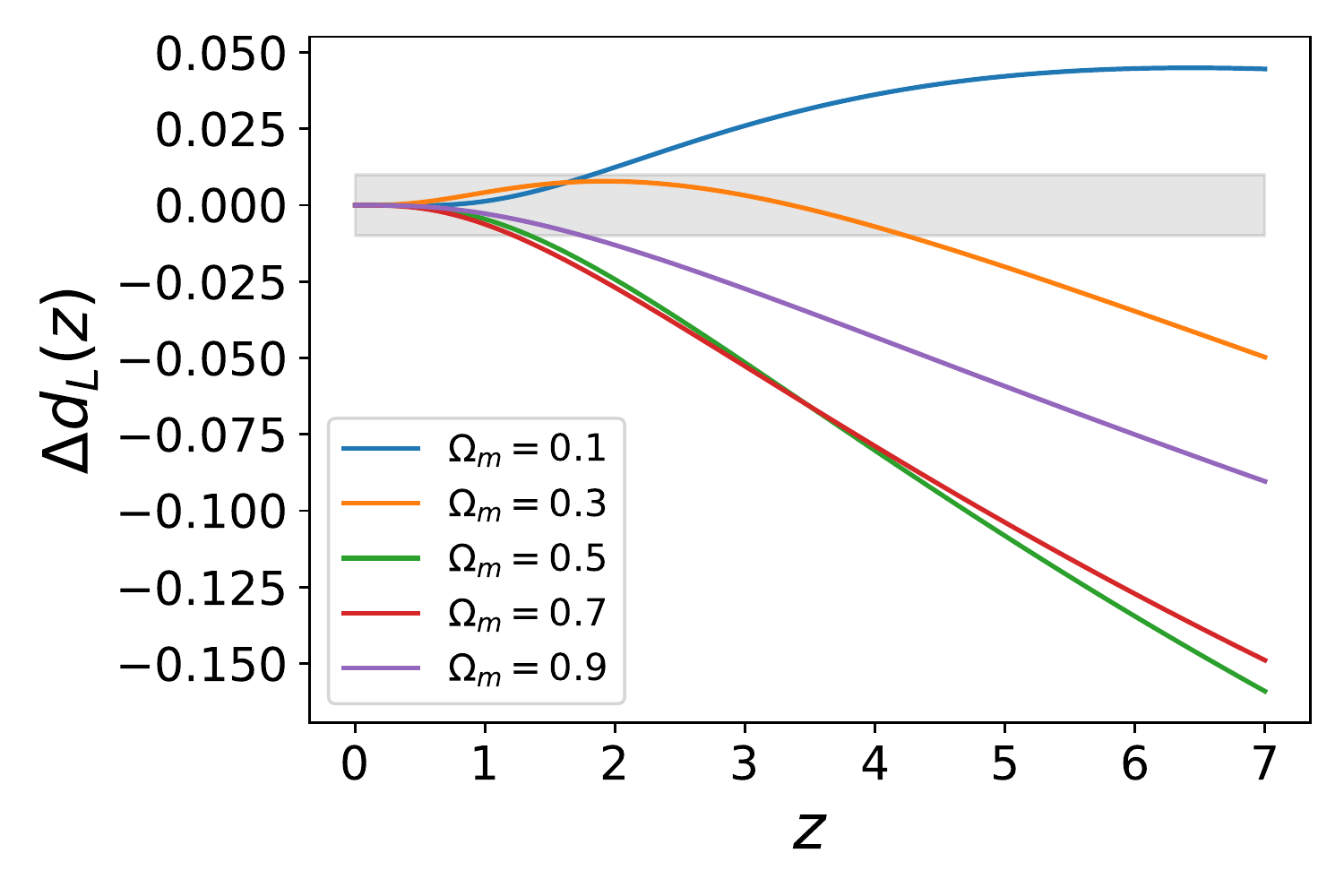}
  \caption{}
\end{subfigure}%
\begin{subfigure}{.35\textwidth}
  \includegraphics[width=55mm]{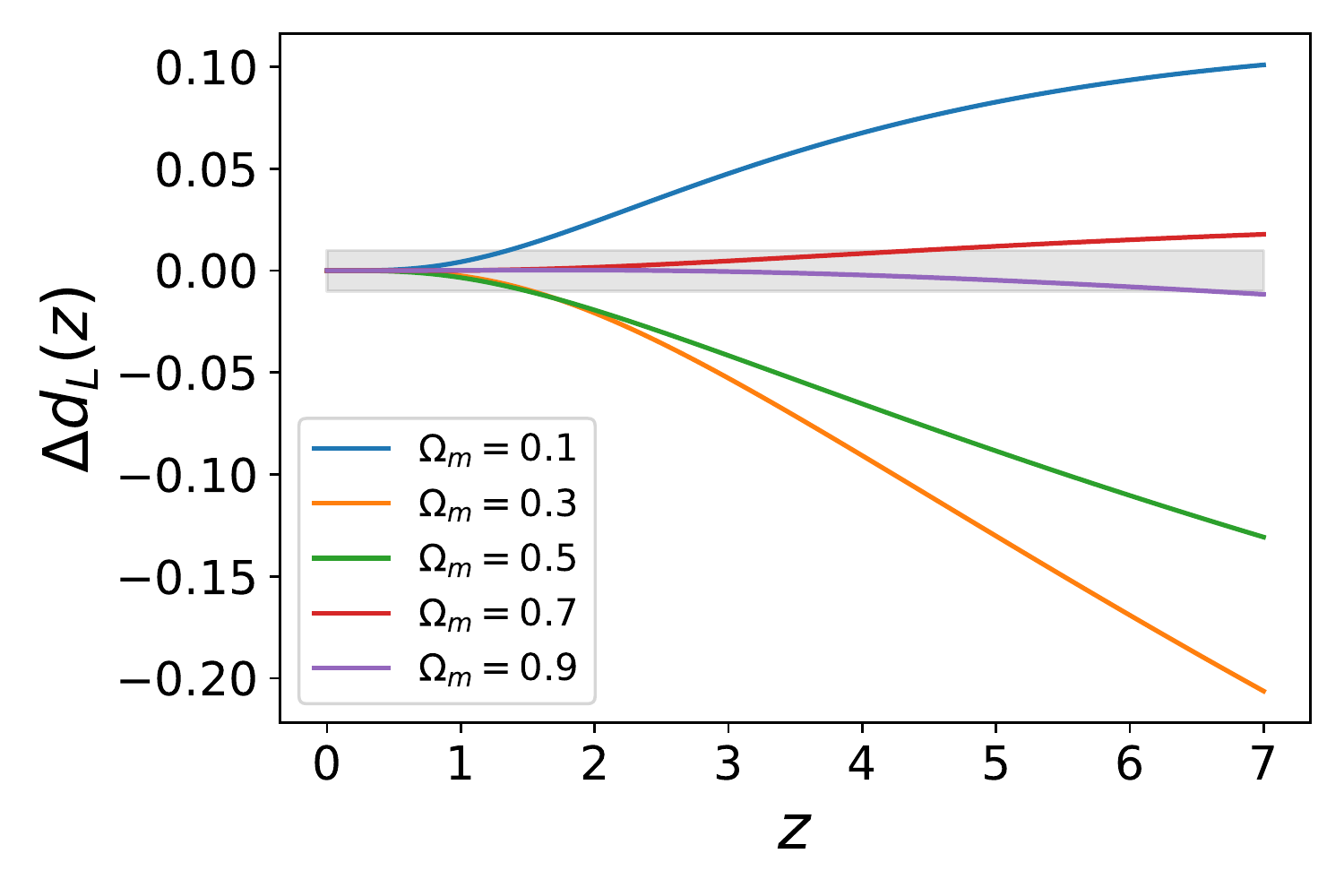}
  \caption{}
\end{subfigure}
\caption{Fractional difference in $d_{L}(z)$ between the $n=3, 4, 5$ polynomials and flat $\Lambda$CDM.}
\label{approx}
\end{figure}

Defining the fractional difference in the luminosity distance, 
\begin{equation}
\Delta d_{L} (z) = \frac{d_{L}^{\textrm{log poly}}(z) - d_{L}^{\Lambda \textrm{CDM}}(z)}{d_{L}^{\Lambda \textrm{CDM}}(z)}, 
\end{equation}
in Fig. \ref{approx} we plot the difference between the $d^{\textrm{log poly}}_{L} (z)$ and $d_{L}^{\Lambda \textrm{CDM}}(z)$ for $0.1 \leq \Omega_m \leq 0.9$ and $n=3$ (third order) \cite{Risaliti:2018reu}, $n=4$ (fourth order) \cite{Lusso:2019akb} and $n=5$ (fifth order) \cite{Lusso:2020pdb}. See  \cite{Yang:2019vgk} for a discussion on $n=4$, but here we illustrate the problem more generally. To remove clutter we only plot five values of $\Omega_m$. As can be seen, deviations typically emerge beyond $ z \sim 1.5-2$, where the fractional difference starts to exceed 1\%, but there are values of $\Omega_m$ that perform better. However, these are merely coincidental. Interestingly, $\Omega_m = 0.7$ and $\Omega_m = 0.9$ both perform better than $\Omega_m = 0.3$ for $n=5$ and the process is largely random. 

It should be stressed that {the above problem arises because the expansions are used outside of the radius of convergence, which is typically $|z|=1$ for cosmological applications. When working outside the radius of convergence, the addition of higher order terms in the expansion does not improve the precision of the expansion.}  {These features are of course expected  \cite{Cattoen:2007sk,Colgain:2021ngq}}. Finally, since the validity of the log polynomial approximation depends on $\Omega_m$, the log polynomial is clearly not ``model-independent".

The key take home for the reader is that Fig. \ref{approx} represents a variable yardstick in the sense that the approximation to flat $\Lambda$CDM, which is the reference model, depends on i) the cosmological parameter $\Omega_m$ and ii) the order $n$ of the expansion. Thus, for different values of $\Omega_m$ and $n$, the  discrepancies or tensions with flat $\Lambda$CDM will vary. Needless to say, employing a variable ``yardstick" runs contrary to best practice in science and should be enough to nullify results quoted in \cite{Risaliti:2018reu, Lusso:2019akb, Lusso:2020pdb}. Fig. \ref{approx} is our main message, but we turn to illustrating the consequences with mock data.

\section*{Illustrative examples}
In order to pinpoint the problem with the log polynomial expansion, we rely upon mock flat $\Lambda$CDM data. For simplicity, we consider the $n=3$ polynomial, since it has the smallest parameter space $(H_0, a_2, a_3)$ and our point is most easily visualised. The lessons for higher-dimensional models are immediate. As is clear from Fig. \ref{approx} (a), the approximations to flat $\Lambda$CDM with $\Omega_m = 0.1$ and $\Omega_m = 0.9$ are the worst, so we focus on these values. In short, we will construct mock data with $H_0 = 70$ km/s/Mpc and $\Omega_m = 0.1$ and $\Omega_m = 0.9$ in order to test whether the log polynomial expansion can recover the value of $\Omega_m$ through a comparison in the $(a_2, a_3)$-plane.\footnote{We take our redshift distributions from \cite{Risaliti:2018reu}. See section 3.1 of  \cite{Yang:2019vgk} for further details.}

First, we recover the cosmological parameters from the mock data through a direct fit of the flat $\Lambda$CDM model to the mock data. The best-fit values inferred from our Markov Chain Monte Carlo (MCMC) chains are shown in Table \ref{table1}. This allows us to quantify the degree of randomness introduced in the mocking procedure\footnote{We have repeated an additional three times and found similar results.} and confirm the data is fully consistent with flat $\Lambda$CDM, in line with expectations. 
\begin{table}[htb]
\centering
\begin{tabular}{ccc}
\hline
\hline
\rule{0pt}{3ex} mock data & $H_0$ (km/s/Mpc) & $\Omega_m$ \\
\hline 
\rule{0pt}{3ex}  $\Omega_m = 0.1$ & $70.053^{+0.143}_{-0.142}$ & $0.102^{+0.003}_{-0.003}$  \\
\rule{0pt}{3ex}  $\Omega_m = 0.9$ & $70.163^{+0.212}_{-0.209}$ & $0.898^{+0.013}_{-0.013}$  \\
\hline 
\end{tabular}
\caption{Best-fit values of $H_0, \Omega_m$ from the mock data.} \label{table1}
\end{table} 

We next repeat the MCMC analysis for the log polynomial expansion and from the chain, it is straightforward to identify confidence interval ellipses. In order to identify the flat $\Lambda$CDM model in the $(a_2, a_3)$-plane, one eliminates $\Omega_m$ from the first two entries in (\ref{equation}) to get the equation \cite{Risaliti:2018reu}: 

\be
\label{curve}
a_3 = \frac{1}{3} (6 a_2^2 - 10 a_2 \ln(10) + 5 \ln(10)^2). 
\ee

As is clear from Fig. \ref{plots}, the log polynomial expansion (\ref{dl}) struggles to recover the cosmological parameters of the underlying data. Worse still, it fails to identify the data as that of the flat $\Lambda$CDM model by an excess of $8\, \sigma$ for the $\Omega_m = 0.1$ mock and $6 \, \sigma$ for the $\Omega_m = 0.9$ mock. It is worth noting that the blue curve corresponding to flat $\Lambda$CDM is the same in both plots, but the highlighted segments in red correspond to the best-fit values of the flat $\Lambda$CDM model recorded in Table \ref{table1}. Interestingly, one can see that the deviation is perfectly correlated with Fig. \ref{approx}, where it is evident that the $\Omega_m = 0.1$ approximation performs worse than $\Omega_m = 0.9$. This is indeed reflected in the magnitude of the deviation from the flat $\Lambda$CDM model. 

\begin{figure}[htb]
\centering
\begin{subfigure}{.5\textwidth}
  \centering
  \includegraphics[width=85mm]{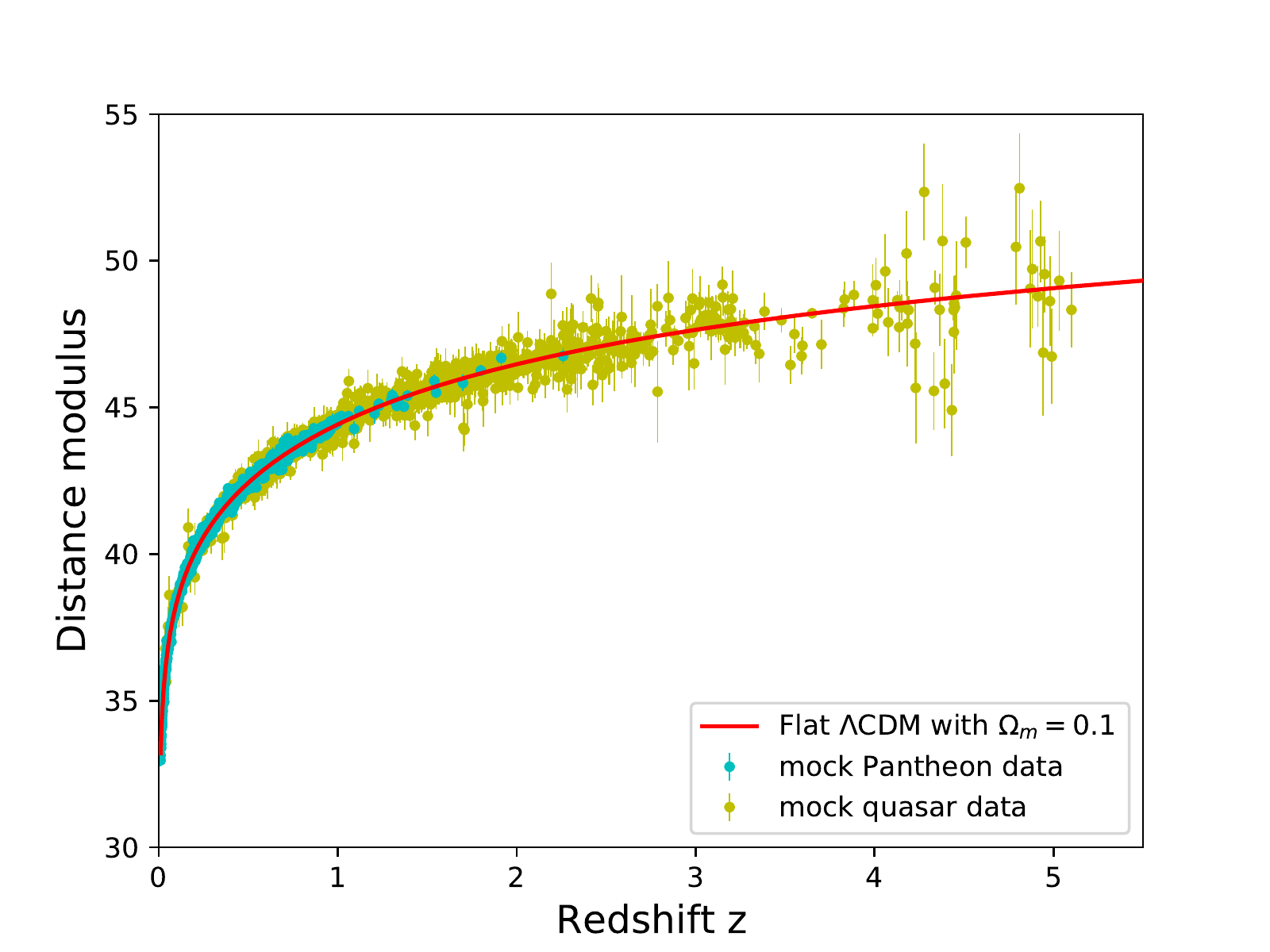}
\end{subfigure}%
\begin{subfigure}{.5\textwidth}
  \centering
  \includegraphics[width=85mm]{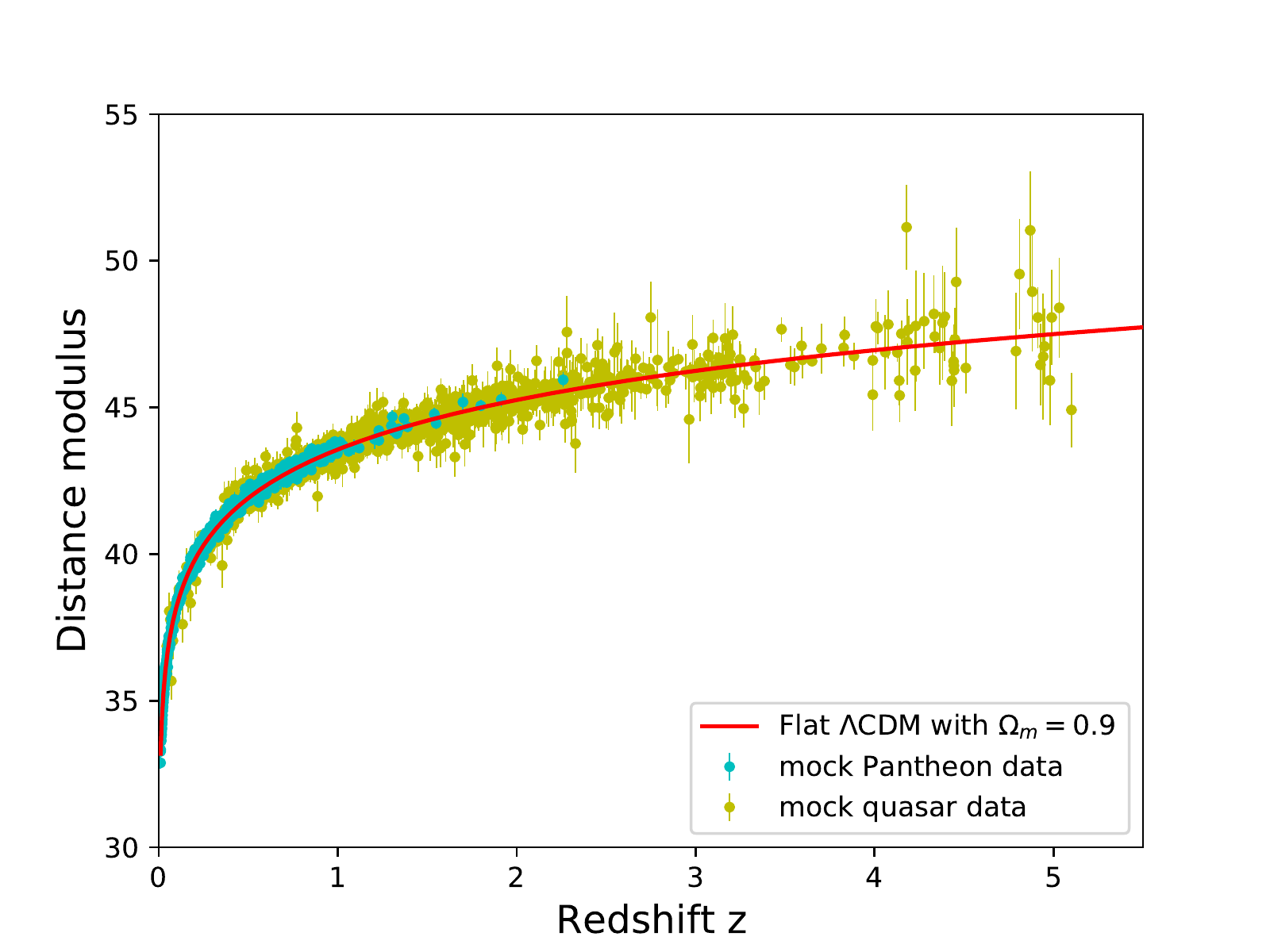}
\end{subfigure}
\caption{Mock flat $\Lambda$CDM data for $\Omega_m = 0.1$ (left) and $\Omega_m = 0.9$ (right). }
\label{mockdata}
\end{figure}

\begin{figure}[htb]
\centering
\begin{subfigure}{.5\textwidth}
  \centering
  \includegraphics[width=80mm]{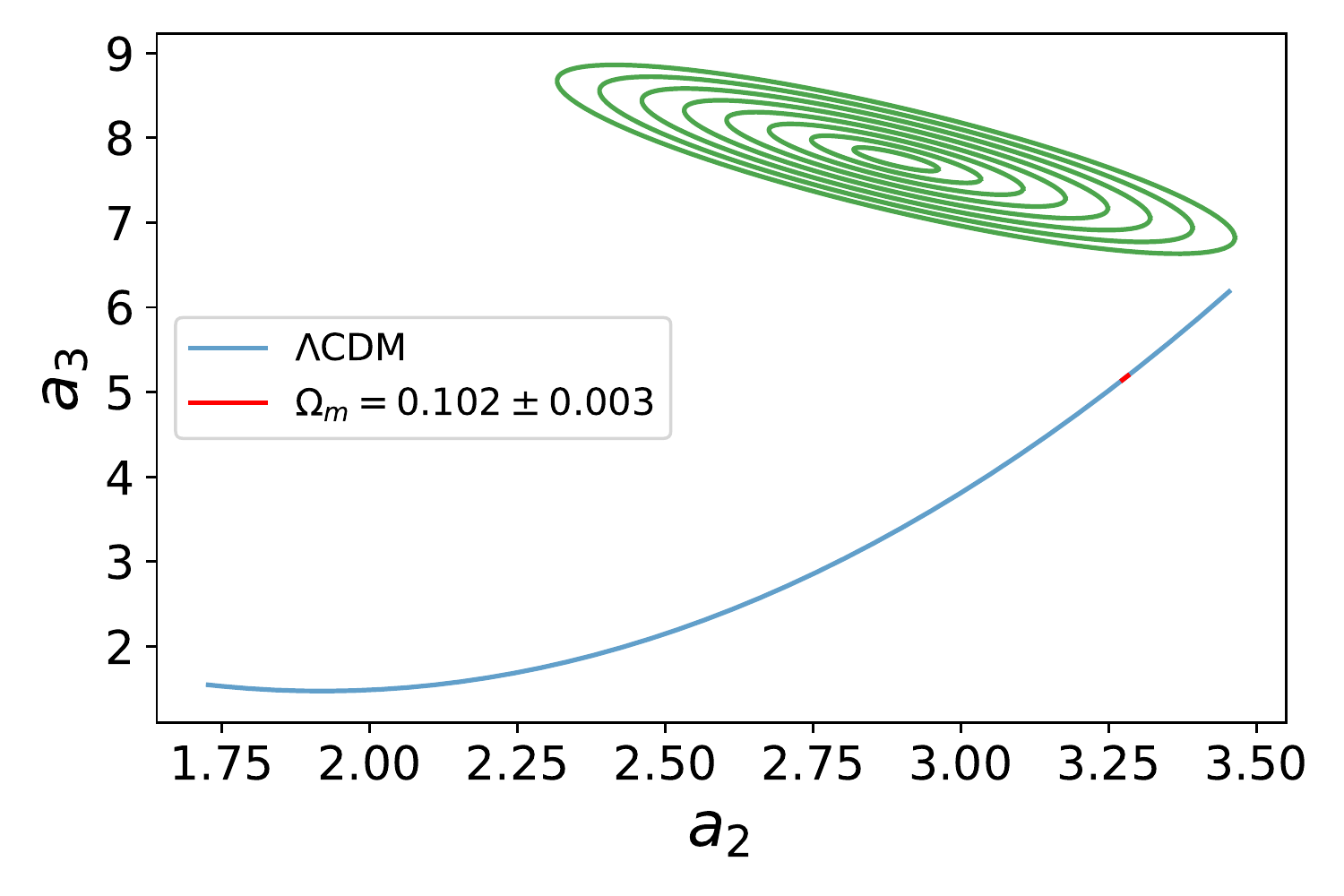}
\end{subfigure}%
\begin{subfigure}{.5\textwidth}
  \centering
  \includegraphics[width=80mm]{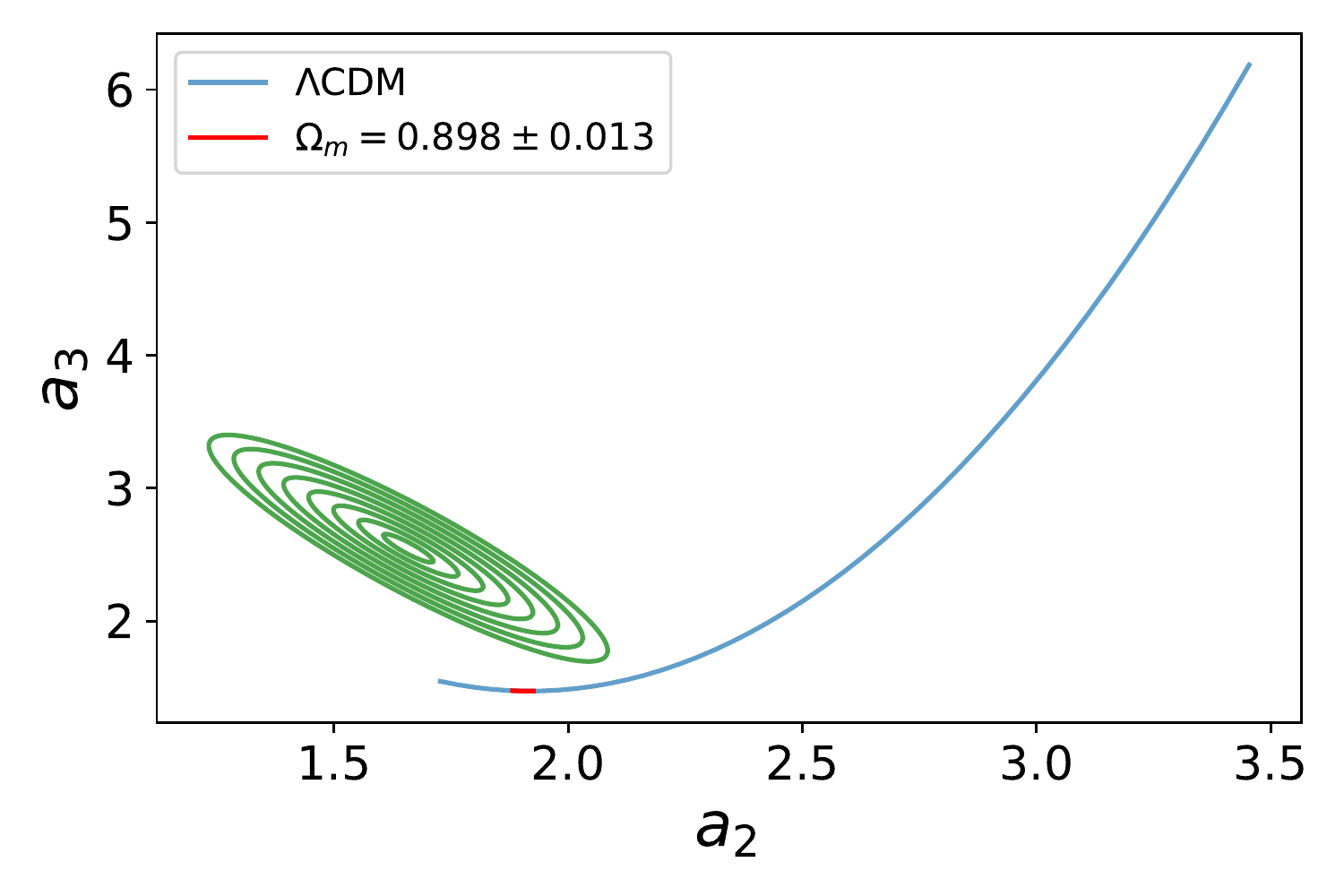}
\end{subfigure}
\caption{Confidence ellipses corresponding to best-fits of the log polynomial with $n=3$ to mock flat $\Lambda$CDM data with $\Omega_m = 0.1$ (left) and $\Omega_m = 0.9$ (right).  The blue curve denotes the flat $\Lambda$CDM family for any value of $\Omega_m$ (\ref{curve}), while the best-fit values from Table \ref{table1} are highlighted by the red segments of the curve.}
\label{plots}
\end{figure}

Recently, prompted by comments in \cite{Yang:2019vgk} and here, an attempt has been made to fix the ``variable yardstick" \cite{Bargiacchi:2021fow}. The approach advocated is to employ an ``orthogonal logarithmic polynomial expansion'':
\be
\label{DL}
D^{\textrm{log poly}}_{L} (z) = \frac{c  \ln (10)}{H_0} \left[ \log_{10} (1+z) + a_2 \log^2 _{10} (1+z)  + a_3 [k_{32}  \log^2_{10} (1+z)^2 + \log^3_{10} (1+z)] \right],  
\ee
where an additional \textit{constant} parameter $k_{32}$ is included. Note, we have simply restricted our attention to the third-order polynomial \cite{Risaliti:2018reu}, which is sufficient to make a point. While one can determine $k_{32}$ as outlined in \cite{Bargiacchi:2021fow}, this term is motivated on the grounds that it removes correlations in the $(a_2, a_3)$ plane. This leads to a change in expressions (\ref{curve}) used to make contact with flat $\Lambda$CDM  \cite{Bargiacchi:2021fow}: 
\be
\label{ka2a3}
a_2 = - k_{32} a_3 + \ln (10) \left( \frac{3}{2} - \frac{3}{4} \Omega_{m} \right), \quad a_3 = \ln^2(10) \left( \frac{9}{8} \Omega_{m}^2 - 2 \Omega_{m} + \frac{7}{6} \right).
\ee

\begin{figure}[htb]
\centering
\begin{subfigure}{.5\textwidth}
  \centering
  \includegraphics[width=80mm]{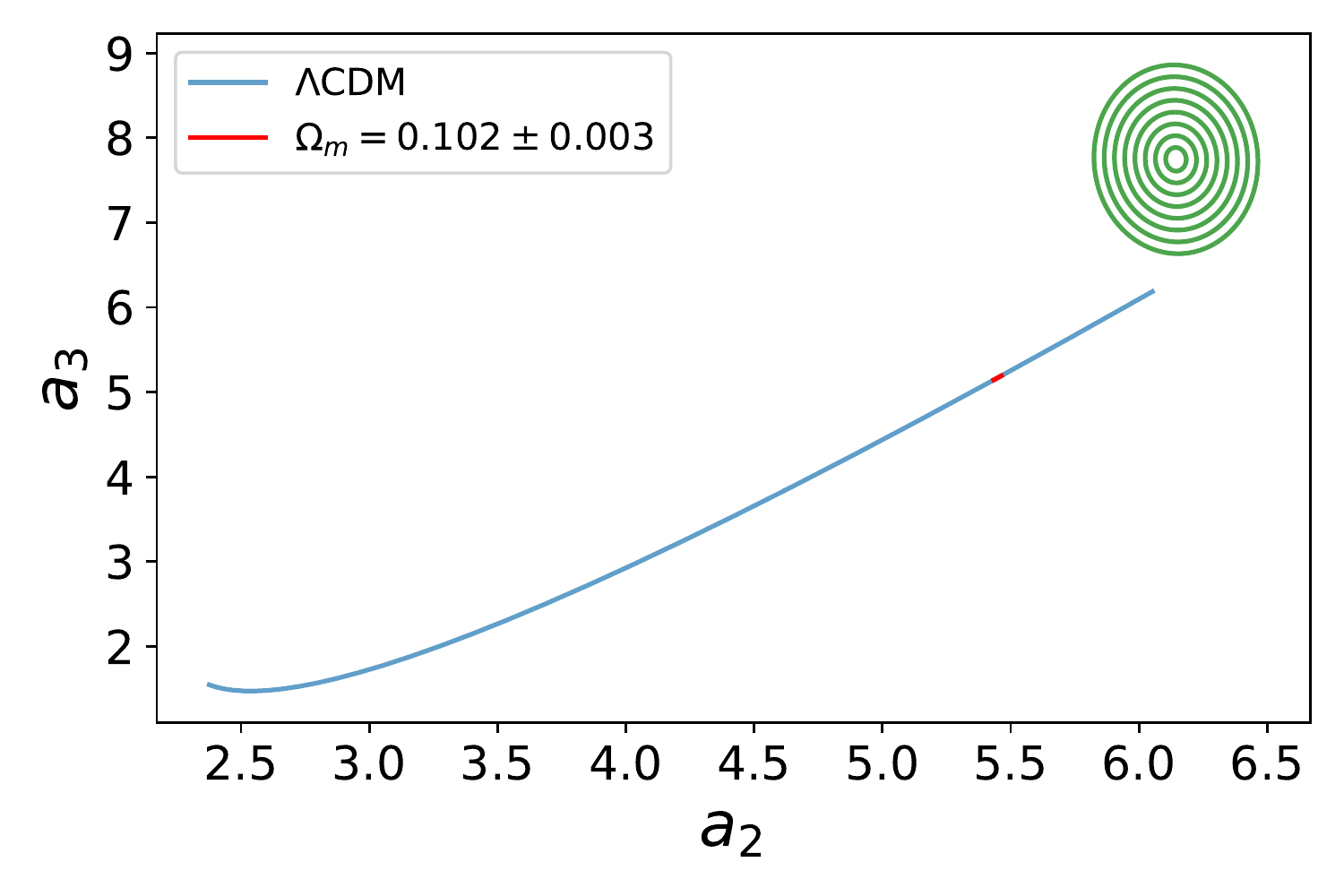}
\end{subfigure}%
\begin{subfigure}{.5\textwidth}
  \centering
  \includegraphics[width=80mm]{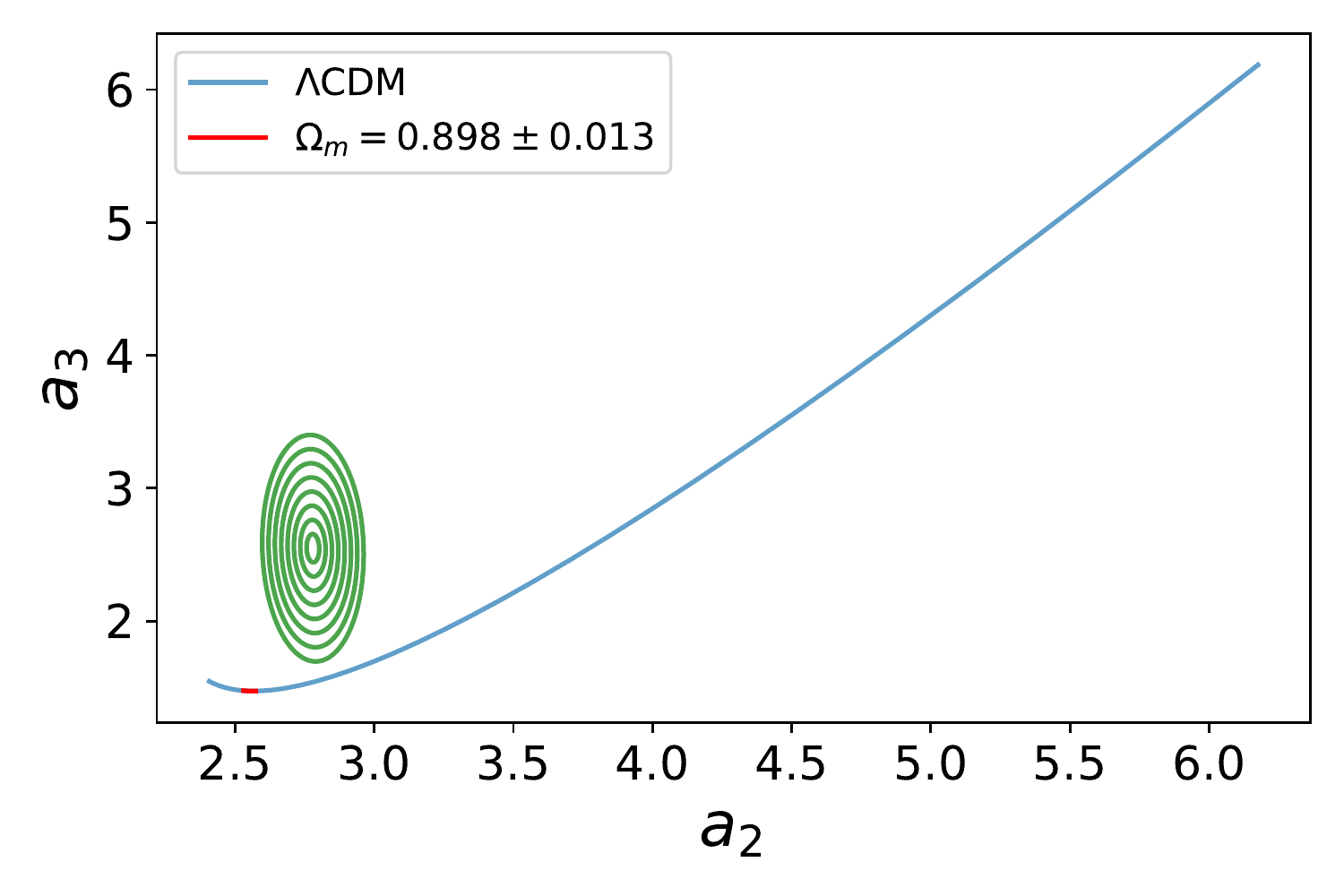}
\end{subfigure}
\caption{Same as Fig. \ref{plots} but with the parameter $k_{32}$ included.}
\label{tilted_plots}
\end{figure}

Note that the third-order term in the orthogonal log polynomial expansion (\ref{DL}) is unchanged from (\ref{dl}). Thus, whether we fit (\ref{dl}) or (\ref{DL}), the best-fit $a_3$ term is the same. To get the new $a_2$ one just redefines $a_2 \rightarrow a_2 - k_{32} a_{3}$. This will affect our ellipses in Fig.  \ref{plots}, and since we know the outcome, i. e. $a_2$ and $a_3$ should be uncorrelated, we can easily infer it via trial and error. The result is shown in Fig. \ref{tilted_plots}. Evidently, our ellipses are not tilted, the flat $\Lambda$CDM curve has also changed and moved to higher $a_2$, but the key message is that the red segments are no closer to the centre of the ellipses.\footnote{For our mocks, we found the value $k_{32} \approx -0.43$.}

\section*{Discussion}
The log polynomial expansion (\ref{dl}) is the basis of a strong claim \cite{Risaliti:2018reu, Lusso:2019akb} in the literature that there is a ``$\sim 4 \sigma$ deviation from the flat $\Lambda$CDM model" when fitted to a compilation of supernovae \cite{Scolnic:2017caz}, quasar and gamma-ray burst data  \cite{Demianski:2016zxi, Demianski:2016dsa}. The key point of Fig. \ref{approx} is that the approximation to the exact luminosity distance of flat $\Lambda$CDM typically only holds up to redshifts  $z \sim 1.5 - 2$. Moreover, in progressing from $n=3$ to $n=5$, there is no guarantee that the approximation improves, and even if the log polynomial performs well for a certain $\Omega_m$, this is purely a coincidence. Therefore, (\ref{dl}) represents a ``variable yardstick". 

To better illustrate the problem, we focused on the $n=3$ case and showed that the log polynomial fails to identify the flat $\Lambda$CDM model, even in cases where the data is fully consistent with the flat $\Lambda$CDM model. Thus, we are left with ``phantom tensions", which are simply an artifact of the expansion. Interestingly, the degree of deviation can be correlated with Fig. \ref{approx}, whereby the $\Omega_m = 0.1$ approximation is worse than $\Omega_m = 0.9$, and this is reflected in the number of ellipses in Fig. \ref{plots}.

Given these observations, one starts to view Fig. 3 of \cite{Risaliti:2018reu} and Fig. 3 \& 4 of \cite{Lusso:2019akb} in a new light. It is telling that the tensions are less for intermediate values of $\Omega_m$, but greatly increase for smaller and higher values. As our analysis demonstrates here, this follows from a breakdown of the log polynomial expansion. 

Overall, great care should be taken with cosmography at high redshift and one typically has to work hard to ensure that polynomial expansions are not impacted by a breakdown of the model. The recent attempt \cite{Bargiacchi:2021fow} to fix the problem is intriguing. {As our analysis above shows, this does not address the varying ``yardstick" issue} (recall Fig. \ref{approx}). What we can say unequivocally is that the new  methodology does not improve the tensions evident in our mock analysis. One may find special circumstances where it works, but mathematics is not on your side.

\section*{Acknowledgments}
We thank S. Appleby, S. Capozziello, E. Di Valentino, O. Luongo and A. Riess for discussion. This work was supported in part by the Korea Ministry of Science, ICT \& Future Planning, Gyeongsangbuk-do and Pohang City. MS acknowledges the support from the grants JSPS KAKENHI Nos. 19H01895 and 20H04727. MMShJ acknowledges the support by INSF grant No 950124 and Saramadan grant No. ISEF/M/99131.



\end{document}